\documentclass[12pt]{article}
\usepackage[
top = 2.5cm, 
bottom = 2.5cm,  
left = 2.55cm,  
right = 2.55cm]{geometry}

\usepackage{bbm}
\usepackage{overpic}
\usepackage{subfigure}
\usepackage{caption}
\usepackage{mathtools}
\usepackage{latexsym} 
\usepackage{verbatim}  
\usepackage{tikz}
\usetikzlibrary{matrix}
\usepackage{color}
\usepackage{graphicx,amssymb,amsfonts,amsmath,amssymb,amscd,amstext, mathrsfs}
\usepackage{graphicx}
\usepackage{dsfont}
\usepackage{xcolor}
\usepackage{amsmath}
\usepackage{empheq}
\usepackage{cite}
\usepackage{caption}

\numberwithin{equation}{section}

\newlength\dlf

\newcommand{\bw}{\begin{widetext}}
\newcommand{\ew}{\end{widetext}}
\newcommand{\bea}{\begin{eqnarray}}
\newcommand{\eea}{\end{eqnarray}}
\newcommand{\be}{\begin{equation}}
\newcommand{\ee}{\end{equation}}

\renewcommand{\bar}[1]{\overline{#1}}

\renewcommand{\hat}[1]{\widehat{#1}}

\renewcommand{\cal}{\mathcal}

\DeclareFontShape{OT1}{cmr}{mx}{n}{<->cmr10}{}
\newcommand{\titlefont}{\fontseries{mx}\selectfont}

\def\frac#1#2{{#1\over #2}}

\usepackage{jheppub}

\begin{document}

\begin{titlepage}

\begin{flushright} 
\end{flushright}

\begin{center} 

\vspace{0.35cm}

{\fontsize{20.5pt}{25pt}
{\titlefont 
Improving Modular Bootstrap Bounds with Integrality
}}

\vspace{1.6cm}  

{{A. Liam Fitzpatrick,  Wei Li}}

\vspace{1cm} 

{{\it
Department of Physics, Boston University, 
Boston, MA  02215, USA
}}\\
\end{center}
\vspace{1.5cm}

{\noindent 
We implement methods that efficiently impose integrality -- i.e., the condition that the coefficients of characters in the partition function must be integers -- into numerical modular bootstrap. We demonstrate the method with a number of examples where it can be used to strengthen modular bootstrap results.  First, we show that,  with a mild extra assumption, imposing integrality improves the bound on the maximal allowed gap in dimensions of operators in theories with a $U(1)^c$ symmetry at $c=3$, and reduces it to the value  saturated by the $SU(4)_1$ WZW model point of $c=3$ Narain lattices moduli space. 
Second, we show that our method  can be used to eliminate all but a discrete set of points saturating the bound from previous Virasoro modular bootstrap results. Finally, when central charge is close to $1$, we can slightly improve the upper bound on the  scaling dimension gap.
}

\end{titlepage}

\tableofcontents

\newpage

\newpage
\section{Introduction} 
Unitary Conformal  Field Theories (CFTs) must satisfy a number of highly restrictive conditions, which can be used to derive bounds on their spectrum and OPE coefficients.   Two particularly powerful conditions are crossing symmetry of correlators, and, in the case of 2d CFTs, modular invariance of the torus partition function. Combined with the constraints of conformal symmetry and unitarity, they lead to the conformal bootstrap and the modular bootstrap, respectively \cite{Rattazzi:2008pe,Hellerman:2009bu}.  These two conditions are structurally equivalent, so that usually analytic or numeric methods developed to study one of them can be applied to the other.  One major way in which they differ is that the partition function is at heart a sum over states, and consequently the degeneracies of operators that appear in the partition function must be integers.  This additional `integrality' constraint is harder to implement in practice, but despite this a few encouraging results have appeared in the literature using integrality to constrain the space of 2d CFTs \cite{Benjamin:2020zbs,Kaidi:2020ecu}.

%

In this paper, we will develop and implement new methods to incorporate the constraint of intergrality into the numerically modular bootstrap. In some cases, we will focus on the integer constraint  on the number of specific operators in the partition function, whereas in other cases we will focus on bound the total degeneracy number of operators with dimensions in a interval $\Delta_a\leq\Delta\leq\Delta_b$. A key part of the strategy will be to obtain two-sided bounds  (i.e., an upper- and a lower-bound) on these degeneracies, in which case we can impose the constraint that \emph{there must be an integer inside the upper and lower bound}. We will describe these methods in detail in Sec.\ref{sec:method}. Starting from Sec.\ref{sec:c3Narain}, we will apply these methods to various examples.  In some cases, we do not assume any symmetry beyond Virasoro symmetry, whereas in other we assume an additional $U(1)^c$ current algebra. Our major results are as follows:
\newline
\newline
\emph{Bootstrapping the $c=3$ Narain CFT}
\newline
\newline
A frequent target for modular bootstrap bounds is to determine the largest possible allowed gap in dimensions of operators between the identity (i.e., vacuum state) and the next lowest-dimension operator.  It is an open question whether or not the optimal bound in the case of two-dimensional CFTs with $U(1)^c\times U(1)^c$ algebra (for integer $c$) can be saturated by  Narain lattices CFTs. Previous modular bootstrap analyses \cite{Collier:2016cls,Hartman:2019pcd,Afkhami-Jeddi:2020ezh} found that this question could be answered positively at some small values of $c$, specifically at $c=1,2,4$.   A natural question is whether the optimal bound can be improved so that it is also saturated by Narain CFTs at $c=3$, or whether there exist unitary $c=3$ CFTs with a larger gap.  
%
  In Sec.\ref{sec:c3Narain} we will show that the condition of integrality, together with a mild additional assumption on the spectrum, implies that the optimal bound on the gap is indeed saturated by a $c=3$ Narain CFT.
\newline
\newline
\emph{Eliminating Extremal Solutions \& Lowering the Spectrual Gap In Virasoro Bootstrap}
\newline
\newline
The authors of \cite{Collier:2016cls} found that, using only Virasoro symmetry, the optimal spectral gap  converged to $\Delta_{\text{mod}}=\frac{c}{6}+\frac{1}{3}$.  Using the extremal function method \cite{El-Showk:2012vjm,El-Showk:2016mxr}, this means that at the limit of the bound, there is a family of modular invariant partition functions with positive coefficients for each Virasoro character in the partition function. However, only four values of $c$ produced partition functions which corresponded to known CFTs \cite{Collier:2016cls}. In Sec.\ref{sec:extremal}, we use integrality to show that no other points can correspond to true CFTs.  More precisely, we take the spectrum given by the extremal functional method, and derive two-sided bounds on the degeneracies of states for several operators in the spectrum.  The four values of $c$ identified in \cite{Collier:2016cls} are the only four values where all two-sided bounds contain integers.  

Finally in Sec.\ref{sec:cnear1}, we analyze the regime of $c$ close to 1 from above, and show that one can obtain a slight improvement on the bound on the gap itself by imposing integrality.

\section{Methods}\label{sec:method}
\subsection{Setup}
\label{sec:setup}
We begin by considering the partition function $Z(\tau,\bar{\tau})$ of a $2d$ CFT on a torus. All modular transformations can be generated from the $S$ and $T$-transformations:
\begin{align}\label{eqn:ST}
&\text{S-transform:~}Z(\tau,\bar{\tau})=Z\Big(-\frac{1}{\tau},-\frac{1}{\bar{\tau}}\Big), \notag\\
&\text{T-transform:~}Z(\tau,\bar{\tau})=Z(\tau+1,\bar{\tau}+1) .
\end{align}
The partition function can moreover be decomposed into a sum over Virasoro characters with non-negative integer coefficients $n_{h,\bar{h}}$ that count the degeneracy of states with weight $h,\bar{h}$:
\begin{equation}
Z(\tau,\bar{\tau})=\chi_{vac}(\tau)\bar{\chi}_{vac}(\bar{\tau})+\sum_{h,\bar{h}}n_{h,\bar{h}}\chi_h(\tau)\bar{\chi}_h(\bar{\tau})
\end{equation}
The vacuum $\chi_{vac}$ and non-degenerate Virasoro character $\chi_h$ are given by
\begin{equation}
\chi_{vac}(\tau)=\frac{q^{-\frac{c-1}{24}}}{\eta(\tau)}(1-q),\quad\chi_h(\tau)=\frac{q^{h-\frac{c-1}{24}}}{\eta(\tau)},
\end{equation}
where $q=e^{2\pi i\tau}$ and $\eta(\tau)$ is the Dedekind eta function. For the rest of paper, we will further assume the partition function is parity-invariant $Z(\tau,\bar{\tau})=Z(\bar{\tau},\tau)$. In this case we can label the Virasoro primaries by their scaling dimension $\Delta=h+\bar{h}$ and their spin $\ell=|h-\bar{h}|$. Then the character decomposition becomes 
\begin{equation}\label{eqn:unitarity}
Z(\tau,\bar{\tau})=\chi_{vac}(\tau)\bar{\chi}_{vac}(\bar{\tau})+\sum_{\Delta,\ell}n_{\Delta,\ell}\Big[\chi_{\frac{\Delta+\ell}{2}}(\tau)\bar{\chi}_{\frac{\Delta-\ell}{2}}(\bar{\tau})+\chi_{\frac{\Delta-\ell}{2}}(\tau)\bar{\chi}_{\frac{\Delta+\ell}{2}}(\bar{\tau})\Big] .
\end{equation}
As first introduced in \cite{Friedan:2013cba}, we can define reduced partition function $\hat{Z}(\tau,\bar{\tau})$ to get rid of the Dedekind eta function,
\begin{equation}
\hat{Z}(\tau,\bar{\tau})=|\tau|^{\frac{1}{2}}|\eta(\tau)|^2Z(\tau,\bar{\tau})
\end{equation}
The extra factor of $|\tau|^{\frac{1}{2}}$ is present because the combination  $|\tau|^{\frac{1}{2}}|\eta(\tau)|^2$ is invariant under $\tau\to-1/\tau$ and consequently so is $\hat{Z}(\tau, \bar{\tau})$.

The $T$-transformation just tells us  that spin $\ell$ must be an integer \eqref{eqn:unitarity}, so we only have to consider the $S$-transformation. Combining \eqref{eqn:unitarity} and \eqref{eqn:ST} we have,
\begin{equation}
\hat{F}_{\text{vac}}(\tau,\bar{\tau})+\sum_{\Delta,\ell}n_{\Delta,\ell}\hat{F}_{\Delta,\ell}(\tau,\bar{\tau})=0,
\label{eq:ModularBootstrapEq}
\end{equation}
where
\begin{align}\label{eqn:modularcrossing}
&\hat{F}_{\text{vac}}(\tau,\bar{\tau})=\hat{\chi}_{\text{vac}}(\tau)\hat{\bar{\chi}}_{\text{vac}}(\bar{\tau})-(\tau\to-\frac{1}{\tau},\bar{\tau}\to-\frac{1}{\bar{\tau}}),\notag\\
&\hat{F}_{\Delta,\ell}(\tau,\bar{\tau})=\hat{\chi}_{\frac{\Delta+\ell}{2}}(\tau)\hat{\bar{\chi}}_{\frac{\Delta-\ell}{2}}(\bar{\tau})+\hat{\chi}_{\frac{\Delta-\ell}{2}}(\tau)\hat{\bar{\chi}}_{\frac{\Delta+\ell}{2}}(\bar{\tau})-(\tau\to-\frac{1}{\tau},\bar{\tau}\to-\frac{1}{\bar{\tau}}).
\end{align}
The idea of the modular bootstrap is 
to find a linear functional $\omega[\hat{F}_{\Delta,\ell}(\tau,\bar{\tau})]=\omega(\Delta,\ell)$ such that
\begin{align}
&\omega[\hat{F}_{\text{vac}}]\geq0,\notag\\
&\omega(\Delta,0)\geq0,\quad\text{For~}\Delta\geq\Delta^{*},\notag\\
&\omega(\Delta,\ell)\geq0,\quad\text{For~}\Delta\geq\ell\text{~and~}\ell\geq1 .
\end{align}
This linear functional $\omega$  acting on (\ref{eq:ModularBootstrapEq}) has a positive contribution from $\omega[\hat{F}_{\text{vac}}]$ that must be canceled by negative contributions, which can only come from scalar states with dimension lower than $\Delta_*$.
In other words, the scalar gap must be smaller than $\Delta^{*}$. To implement semi-definite programming\cite{Simmons-Duffin:2015qma,Landry:2019qug} on the modular crossing equation, we will apply the following linear functionals $\omega^{(m,n)}$ as introduced in \cite{Afkhami-Jeddi:2021iuw,Afkhami-Jeddi:2019zci},
\begin{align}
&\omega^{(k,\bar{k})}(\Delta,\ell)=\mathcal{D}_\beta^{k}\mathcal{D}_{\bar{\beta}}^{\bar{k}}\hat{F}_{\Delta,\ell}(i\beta,-i\bar{\beta})\Big|_{\beta=\bar{\beta}=1}\notag\\
&=e^{-4\pi(\Delta-\frac{c-1}{12})}L_{2k-1}^{(-1/2)}\Big[4\pi\Big(\frac{\Delta+\ell}{2}-\frac{c-1}{24}\Big)\Big]L_{2\bar{k}-1}^{(-1/2)}\Big[4\pi\Big(\frac{\Delta-\ell}{2}-\frac{c-1}{24}\Big)\Big]+(k\leftrightarrow\bar{k}),\notag\\
&\omega^{(k,\bar{k})}[\hat{F}_{\text{vac}}] =\omega^{(k,\bar{k})}(0,0)-2\omega^{(k,\bar{k})}(1,1)+\omega^{(k,\bar{k})}(2,0).
\end{align}
where $k+\bar{k}$ are odd integers and $L_p^{(\alpha)}(x)$ are the generalized Laguerre polynomials. Collecting different components of $(k,\bar{k})$ we get a truncated crossing equation:
\begin{equation}\label{eqn:truncatecrossing}
\vec{F}_{\text{vac}}+\sum_{\Delta,\ell} n_{\Delta,\ell}\vec{F}_{\Delta,\ell}=\vec{0},\quad\vec{F}_{\text{vac}}=
\begin{pmatrix}
\omega^{(1,0)}[\hat{F}_{\text{vac}}]\\
\omega^{(2,1)}[\hat{F}_{\text{vac}}]\\
\vdots\\
\omega^{(k,\bar{k})}[\hat{F}_{\text{vac}}]
\end{pmatrix}
,\quad
\quad\vec{F}_{\Delta,\ell}=
\begin{pmatrix}
\omega^{(1,0)}(\Delta,\ell)\\
\omega^{(2,1)}(\Delta,\ell)\\
\vdots\\
\omega^{(k,\bar{k})}(\Delta,\ell)
\end{pmatrix}
\end{equation}
For the rest of the paper, we will use $\Lambda_{\text{max}}$ to denote the maximal derivative order $k+\bar{k}\leq\Lambda_{\text{max}}$ in \eqref{eqn:truncatecrossing}.

In Sec.\ref{sec:c3Narain}, we will also consider the bootstrap 2D CFTs with $U(1)^c\times U(1)^c$ algebra. In that case, the only modification is that we have to change to Virasoro character to $U(1)^c$ character:
\begin{equation}
\chi_{\text{vac}}^{U(1)^c}=\frac{1}{\eta(\tau)^c},\quad\chi_{h}^{U(1)^c}=\frac{q^h}{\eta(\tau)^c} .
\end{equation}
For the linear functionals, we will use the eigenfunctions found in \cite{Afkhami-Jeddi:2020ezh} as our basis,
\begin{equation}
\begin{aligned}
& \omega^{(k,\bar{k})}_{U(1)^c}(\Delta,\ell)=e^{-2\pi\Delta}\bigg(L_k^{(c/2-1)}\Big[4\pi\Big(\frac{\Delta+\ell}{2}\Big)\Big]L_{\bar{k}}^{(c/2-1)}\Big[4\pi\Big(\frac{\Delta-\ell}{2}\Big)\Big]+(k\leftrightarrow\bar{k})\bigg), \\
&\omega^{(k,\bar{k})}_{U(1)^c}[\hat{F}_{\text{vac}}^{U(1)^c}]=\frac{1}{2}\omega^{(k,\bar{k})}_{U(1)^c}(0,0) .
\end{aligned}
\end{equation}
\subsection{Degeneracy bounds on individual operator}\label{sec:individual}
Let us first review the standard method for bounding the degeneracy numbers $n_{\Delta_{\mathcal{O}},\ell_{\mathcal{O}}}$.
 From Sec.\ref{sec:setup}, modular invariance implies the following crossing equation:

\begin{equation}\label{eqn:crossing}
\vec{F}_{\text{vac}}+n_{\Delta_{\mathcal{O}},\ell_{\mathcal{O}}}\vec{F}_{\Delta_{\mathcal{O}},\ell_{\mathcal{O}}}+\sum_{\{\Delta,\ell\}\in\mathcal{S}}n_{\Delta,\ell}\vec{F}_{\Delta,\ell}=\vec{0}
\end{equation}
where  $\mathcal{S}$ contains all the operators (excluding the vacuum and $\mathcal{O}$)  in the CFT . If there exist a vector $\vec{\alpha}$ satisfying the condition
\begin{equation}\label{eqn:con1}
\text{For all}~\{\Delta,\ell\}\in\mathcal{S}
,\quad\vec{\alpha}\cdot\vec{F}_{\Delta,\ell}\geq0,
\end{equation}
then we can act with $\vec{\alpha}$ on both sides of \eqref{eqn:crossing}, which produces a upper bound or lower bound of $n_{\Delta_{\mathcal{O}},\ell_{\mathcal{O}}}$ depending on the sign of $\vec{\alpha}\cdot\vec{F}_{\Delta_{\mathcal{O}},\ell_{\mathcal{O}}}$. For convenience, we choose $\vec{\alpha}\cdot\vec{F}_{\Delta_{\mathcal{O}},\ell_{\mathcal{O}}}=\pm1$, in which case
\begin{equation}\label{eqn:con2}
\vec{\alpha}\cdot\vec{F}_{\Delta_{\mathcal{O}},\ell_{\mathcal{O}}}=1\Rightarrow n_{\Delta_{\mathcal{O}},\ell_{\mathcal{O}}}\leq-\vec{\alpha}\cdot\vec{F}_{\text{vac}},\quad\vec{\alpha}\cdot\vec{F}_{\Delta_{\mathcal{O}},\ell_{\mathcal{O}}}=-1\Rightarrow n_{\Delta_{\mathcal{O}},\ell_{\mathcal{O}}}\geq\vec{\alpha}\cdot\vec{F}_{\text{vac}} .
\end{equation}  
In these two cases, the strongest upper and lower bound are both coming from maximizing $\vec{\alpha}\cdot\vec{F}_{\text{vac}}$.

The exact same technique has been applied to the conformal bootstrap for correlators, where the result is a bound on OPE coefficients \cite{Caracciolo:2009bx}. But in the modular bootstrap case, integrality imposes an additional constraint:
\begin{equation}\label{eqn:con1}
\textbf{The upper and lower bound of $n_{\Delta_{\mathcal{O},\ell_{\mathcal{O}}}}$ must contain a integer}
\end{equation}

One of the first things one might be tempted to try is simply to implement the above strategy for two-sided bounds on $n_{\Delta_{\mathcal{O},\ell_{\mathcal{O}}}}$, and then impose (\ref{eqn:con1}) in order to improve the bound on the gap.  Unfortunately, this strategy fails.  The reason is that, for the optimal gap, $\Delta_{\cal O}$ is at the infimum of the range of dimensions $\Delta$ of operators in $\mathcal{S}$, but $\vec{\alpha} \cdot \vec{F}_{\Delta, \ell} \ge 0$ for $\Delta \in \mathcal{S}$ by construction. Then, by continuity,  $\vec{\alpha} \cdot \vec{F}_{\Delta_{\mathcal{O}}, \ell_{\mathcal{O}}}$ cannot be negative, so we cannot use (\ref{eqn:con2}) to get lower bounds on $n_{\Delta_{\cal O}, \ell_{\cal O}}$.

\begin{figure}
    \centering
    \subfigure[]{\includegraphics[width=0.42\textwidth]{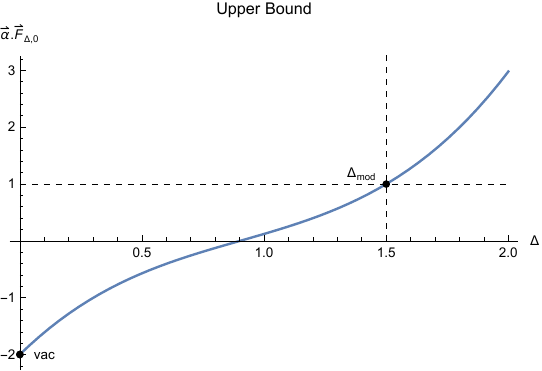}} 
    \quad
    \subfigure[]{\includegraphics[width=0.42\textwidth]{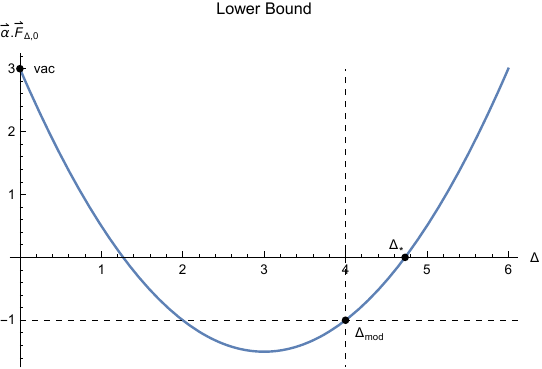}} 
    \caption{(a) Functionals that produces upper bound $n_{\Delta_{\text{mod}},\ell_{\mathcal{O}}=0}$ (b) Functionals that produces lower bound $n_{\Delta_{\text{mod}},\ell_{\mathcal{O}}=0}$, with additional assumption that the next operator has scaling dimension larger than $\Delta_*$}
    \label{fig:oplowerupperbound}
\end{figure}

So if we want to have a two-sided bound on $n_{\Delta_{\text{mod}},\ell_{\mathcal{O}}}$, we will need to make an additional assumption on the spectrum. One option is to impose a second gap $\Delta_*$ in the spin sector $\ell=\ell_{\mathcal{O}}$. Now the operator set $\mathcal{S}$ will become,
\begin{equation}
\mathcal{S}=\{(\Delta,\ell)~|~\text{If }\ell=\ell_{\mathcal{O}},\Delta\geq\Delta_*,~\text{else }\Delta\geq\text{max}(\Delta_{\text{mod}},\ell),~\ell\geq0\}
\end{equation}
In Fig.\ref{fig:oplowerupperbound}, we show two plots for functionals that produces upper and lower bound given by \eqref{eqn:con2}.

Although this method by itself does not help to shrink the previous modular bootstrap bound, it is still a powerful method when we know some information about the next operator. In Sec.\ref{sec:c3Narain} and Sec.\ref{sec:extremal}, we will show two applications of this method.  The first application will be to lower the bootstrap down to a $c=3$ Narain theory under the assumption on the second gap in the $\ell=1$ sector. The second application is to rigorously eliminate all but a small set of extremal solutions from the Virasoro bootstrap in $1\leq c\leq4$.

\subsection{Degeneracy bounds in a interval}\label{sec:interval}
In this section we provide a method to impose the integrality on the interval of operators, inspired by the method from \cite{Collier:2017shs} for bounds on the CFT spectral density.
Let  $\mathcal{F}_\ell(\Delta_a,\Delta_b)$ be the integrated density of operators in an interval $\Delta\in[\Delta_a,\Delta_b]$:
\begin{equation}\label{eqn:densityinterval}
\mathcal{F}_\ell(\Delta_a,\Delta_b)=\sum_{\Delta_a\leq\Delta\leq\Delta_b}n_{\Delta,\ell} .
\end{equation}
If  we can obtain a two sided bound on $A\leq\mathcal{F}_\ell(\Delta_a,\Delta_b)\leq B$,  then we can impose the additional constraint that  $[A,B]$ must contain an integer. 

Now we introduce our method to get an upper and lower bound on $\mathcal{F}_\ell(\Delta_a,\Delta_b)$ using semi-definite programming\cite{Simmons-Duffin:2015qma,Landry:2019qug}. Combining \eqref{eqn:truncatecrossing} and \eqref{eqn:densityinterval} we obtain the following vector equation,
\begin{equation}\label{eqn:intervaleq}
\begin{pmatrix}
0\\
\vec{F}_{\text{vac}}
\end{pmatrix}
+\sum_{\{\Delta,\ell\}\in\mathcal{S}}n_{\Delta,\ell}
\begin{pmatrix}
\theta(\Delta-\Delta_b)\theta(\Delta_a-\Delta)\delta_{s,\ell}\\
\vec{F}_{\Delta,\ell}
\end{pmatrix}=
\begin{pmatrix}
\mathcal{F}_s(\Delta_a,\Delta_b)\\
\vec{0}
\end{pmatrix}
\end{equation}
where $\theta(x)$ and $\delta_{s,l}$ are Heaviside step function and Kronecker delta function. We will seek a vector $\vec{\beta}$,
\begin{equation}
\vec{\beta}\equiv
\begin{pmatrix}
\pm1\\
\vec{\alpha}
\end{pmatrix},
\end{equation}
 such that,
\begin{equation}
\vec{\beta}\cdot
\begin{pmatrix}
\theta(\Delta-\Delta_b)\theta(\Delta_a-\Delta)\delta_{s,\ell}\\
\vec{F}_{\Delta,\ell}
\end{pmatrix}
\geq0~~\text{for }\{\Delta,\ell\}\in\mathcal{S} .
\end{equation}
Then, acting $\vec{\beta}$ on \eqref{eqn:intervaleq} we have,
\begin{equation}
\pm \mathcal{F}_s(\Delta_a,\Delta_b)=  \vec{\alpha}\cdot\vec{F}_{\text{vac}}+ \sum_{\{\Delta,\ell\}\in\mathcal{S}}(n_{\Delta,\ell})\vec{\beta}\cdot
\begin{pmatrix}
\theta(\Delta-\Delta_b)\theta(\Delta_a-\Delta)\delta_{s,\ell}\\
\vec{F}_{\Delta,\ell}
\end{pmatrix}
\geq\vec{\alpha}\cdot\vec{F}_{\text{vac}}
\end{equation}
%
%
So depending on which sign we choose for the top component of $\vec{\beta}$,  we can produce either an upper or lower bound on $\mathcal{F}_s(\Delta_a,\Delta_b)$.
 The strongest bound in both cases will come from maximizing $\vec{\alpha}\cdot\vec{F}_{\text{vac}}$. Below we show two typical functionals that give rise to upper and lower bound of $\mathcal{F}_s(\Delta_a,\Delta_b)$.

In general, we find that by choosing an appropriate interval $[\Delta_a,\Delta_b]$ and spin sector $s$, one can always find an upper and lower bound of $\mathcal{F}_s(\Delta_a,\Delta_b)$. So unlike in Sec.\ref{sec:individual}, we don't need to put additional assumption of the next operator to get a lower bound. Then the natural question is whether this method can lower the previous bootstrap bound? For the problem of bounding the lowest scaling dimension, we find that the answer is yes, though the actual amount by which we will end up improving the bound in our applications will be modest.  
We will discuss this in more detail in Sec.\ref{sec:cnear1}.

\section{Review of Previous Bootstrap Results}\label{sec:bootresult}
Without assuming any symmetry beyond Virasoro, in \cite{Collier:2016cls} the authors found that for $1\leq c \leq4$, the modular bootstrap bound on the overall spectral gap 
converges to
\begin{equation}\label{eqn:modgap}
\Delta_{\text{mod}}=\frac{c}{6}+\frac{1}{3},\quad c\in[1,4] .
\end{equation}
Moreover, they identified the following extremal solutions in terms of known theories:
\begin{itemize}
\item At $c=1$, the bound is $\Delta_{\text{mod}}=\frac{1}{2}$. The theory saturating this bound is free compact boson at self-dual radius, or equivalently $SU(2)_1$ WZW model. 
\item At $c=2$, the bound is $\Delta_{\text{mod}}=\frac{2}{3}$. The theory saturating this bound is the $SU(3)_1$ WZW model
\item At $c=\frac{14}{5}$, the bound is $\Delta_{\text{mod}}=\frac{4}{5}$. The theory saturating this bound is $(G_2)_1$ WZW model
\item At $c=4$, the bound is $\Delta_{\text{mod}}=1$. The theory saturating this bound is $SO(8)_1$ WZW model
\end{itemize}

\begin{figure}
    \centering
    \subfigure[]{\includegraphics[width=0.42\textwidth]{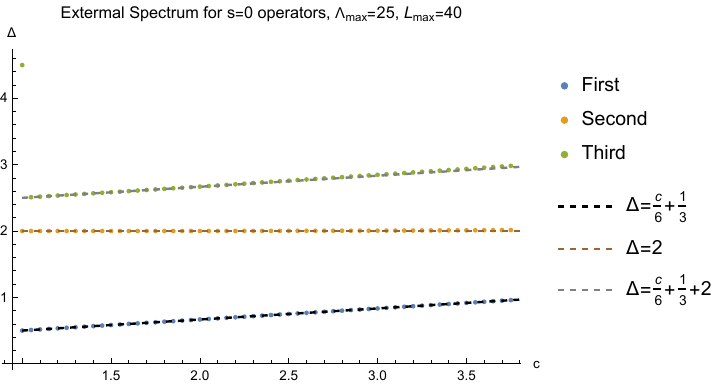}} 
    \subfigure[]{\includegraphics[width=0.42\textwidth]{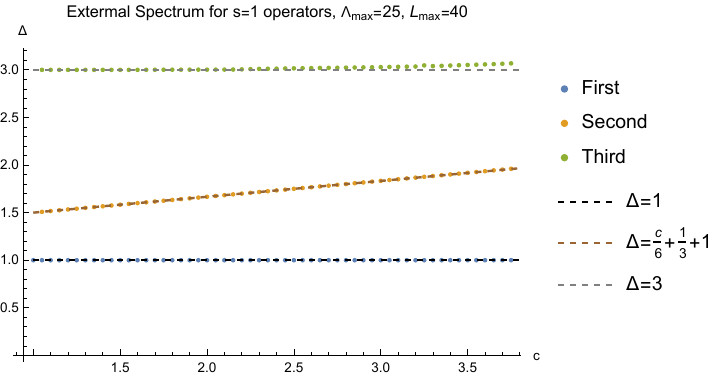}} 
    \subfigure[]{\includegraphics[width=0.42\textwidth]{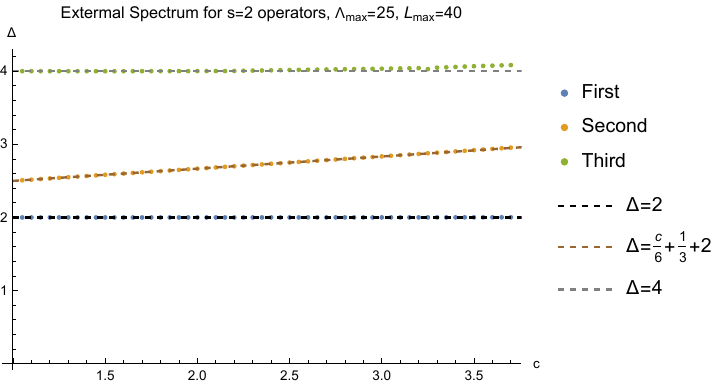}}
    \caption{This plot shows the scaling dimension $\Delta$ of the first three operators in $\ell=0,1,2$. It's extracting from extremal solutions of Virasoro Bootstrap at $\Lambda_{max}=25,L_{max}=40$ (a) $\ell=0$ (b) $\ell=1$ (c) $\ell=2$}
    \label{fig:c1to4spectrum}
\end{figure}

More generally, by looking at the spectrum of the truncated crossing equation at sufficient high derivative order, we observed the spectrum of the extremal solution for the range $1\leq c\leq4$ will have the following pattern:

\begin{align}\label{eqn:spectrum}
&\ell=0,\quad\Delta=\{\Delta_\phi,2,2+\Delta_\phi,4,4+\Delta_\phi,\dots,2n,2n+\Delta_\phi\}\notag\\
&\ell=1,\quad\Delta=\{1,1+\Delta_\phi,3,3+\Delta_\phi,5,5+\Delta_\phi,\dots,2n+1,2n+1+\Delta_\phi\}\notag\\
&\ell=2,\quad\Delta=\{2,2+\Delta_\phi,4,4+\Delta_\phi,\dots,2n,2n+\Delta_\phi\}\notag\\
&\ell=3,\quad\Delta=\{3,3+\Delta_\phi,5,5+\Delta_\phi,\dots,2n+1,2n++1+\Delta_\phi\}\notag\\
&\vdots
\end{align}
where $\Delta_\phi=\frac{c}{6}+\frac{1}{3}$.\footnote{The spectrum suggests that it's likely to be RCFT with two primaries: one is the identity $\mathbf{1}$ and the other one is a scalar $\phi$ with conformal weight $h=\bar{h}=\frac{c}{12}+\frac{1}{6}$. And for these four WZW models shown in the table above, they all have exactly two chiral algebra primaries.} In Fig.\ref{fig:c1to4spectrum}, we show the first few operators from $\Lambda_{max}=25$ extremal solutions to verify this pattern. The convergence of the spectrum means that this is indeed the spectrum of a modular invariant partition function with positive coefficients when decomposed into the characters. But in general we find, most of them don't have integer degeneracy numbers. In Sec.\ref{sec:extremal}, we will obtain two-sided bounds on the degeneracies and show that, assuming the spectrum obeys this pattern, $c=1,2,\frac{14}{5},4$ are the only legitimate solutions. 

We can also consider not just Virasoro symmetry, but assume the extra conserved currents. We focus on the case of 2d CFTs with a $U(1)^c$ chiral algebra. In \cite{Afkhami-Jeddi:2020ezh}, the authors showed
that in this case, the modular bootstrap bounds at $c=1,2,4$, the bound are saturated by Narain lattices CFTs.

\begin{figure}
    \centering
   \includegraphics[width=0.7\textwidth]{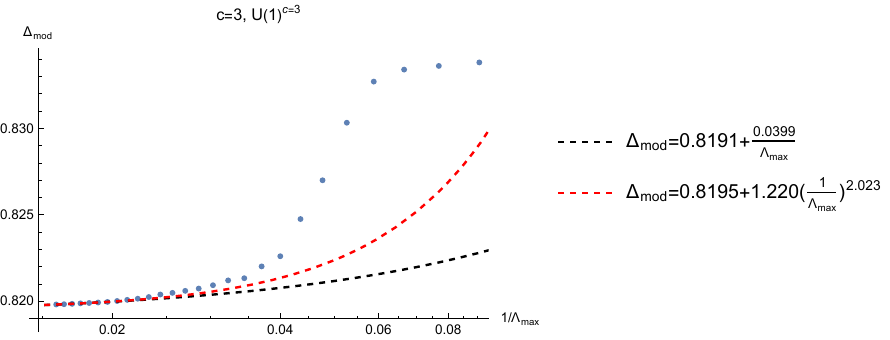}
    \caption{This plot shows the spectral gap from $U(1)^{c=3}$ bootstrap from $\Lambda_{max}=11$ to $\Lambda_{max}=63$. Black and red dashed curves correspond to two different fitting functions: $\Delta_{\text{mod}}=a+\frac{b}{\Lambda_{max}}$ and $\Delta_{\text{mod}}=a+b\Big(\frac{1}{\Lambda_{max}}\Big)^c$}
    \label{fig:c3gap}
\end{figure}

At $c=3$ the authors found the spectral gap $\Delta_{\text{mod}}^{(25)}=0.8227$ while the Narain CFTs at $c=3$ have at most a  spectral gap $\Delta^{\text{Narain}}_{\text{mod}}=\frac{3}{4}$. One might wonder whether the bound rom \cite{Afkhami-Jeddi:2020ezh} could be improved from 0.8227 to 0.75 simply by increasing the truncation on the number of derivatives ($\Lambda_{max}=25$ in their case) in the modular bootstrap analysis. 
%
However, it turns out that this is not the case.  In Fig.\ref{fig:c3gap} we show the modular bootstrap bound up to $\Lambda_{max}=63$. Extrapolating to $\Lambda_{\rm max} = \infty$, we estimate that the infinite truncation modular bootstrap bound is $\Delta_{\text{mod}}^{\Lambda_{max}=\infty}\sim0.819$, which is still not close to $0.75$. In Sec.\ref{sec:c3Narain}, we will argue that the bound can nevertheless be decreased to $0.75$ using integrality together with the assumption that the first two states in the spin-1 sector follow the same pattern as seen in (\ref{fig:c1to4spectrum}) from the extremal spectra.

\section{Bootstrapping the $c=3$ Narain CFT}\label{sec:c3Narain}
In this section we will apply integrality to the bootstrap bound on the spectral gap in theories with a $U(1)^c$ chiral algebra, at $c=3$.  In the previous section, we saw that without any additional assumptions beyond unitarity, the modular bootstrap bound at infinite derivative order appears to asymptote to $\Delta_{\rm mod} \approx 0.819$, whereas the largest possible gap of a $c=3$ Narain CFT is $3/4$.  In this section, we will have to make one additional assumption, which is that the structure seen in the first two states in the spin-1 sector in the extremal spectrum (see (\ref{eqn:spectrum}) from the modular bootstrap bound holds for the true theory that maximizes the spectral gap.  To be precise, we assume that the spin-1 sector has a spin-1 current, followed by a gap to a spin-1 state with dimension $1+\Delta_\phi$, where $\Delta_\phi$ is the spectral gap.  Aside from the fact that this is the extremal spectrum seen from the modular bootstrap, it is also what one would expect if extremal theories have an enhanced symmetry (the presence of at least one extra spin-1 current in the spectrum, which exist in addition to the $U(1)^c$ currents).  We do not, however, rule out the possibility that there might be a $c=3$ theory with a gap between $3/4$ and $0.819$ that has a spin-1 state with dimension between $1$ and $1+\Delta_\phi$. 

\begin{figure}
    \centering
   \includegraphics[width=0.49\textwidth]{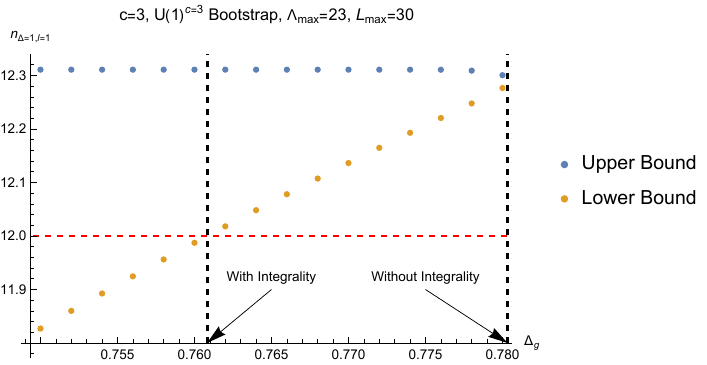}
      \includegraphics[width=0.49\textwidth]{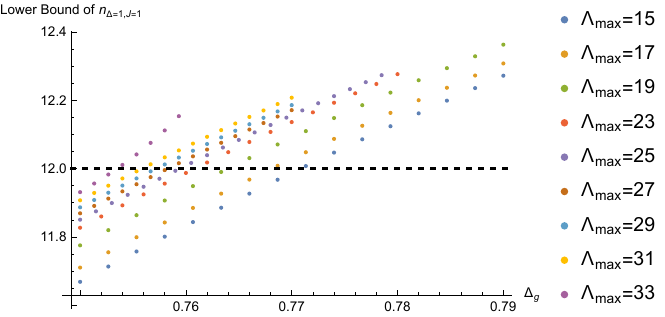}
    \caption{{\it Left:} The upper and lower bounds on $n_{\Delta=1,\ell=1}$ for $U(1)^{c=3}$ bootstrap, at $\Lambda_{max}=23$. {\it Right:} The lower bounds on $n_{\Delta=1,\ell=1}$ for $U(1)^{c=3}$ bootstrap for various derivative orders. }
    \label{fig:c3u1example}
\end{figure}
\begin{figure}
    \centering
   \includegraphics[width=0.7\textwidth]{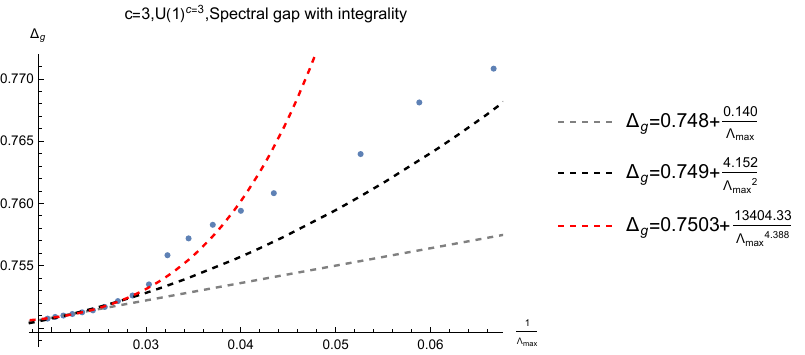}
    \caption{Scalar gap $\Delta_g$ after imposing integrality constraint for derivative orders up to $\Lambda_{max}=51$. Black, gray and red dashed curves are using three different fitting functions $\Delta_g=a+\frac{b}{\Lambda_{max}},\Delta_g=a+\frac{b}{\Lambda_{max}^2},\Delta_g=a+\frac{b}{\Lambda_{max}^c}$ respectively}
    \label{fig:c3u1gap}
\end{figure}

Nevertheless, with this extra assumption, we will find that imposing integrality on the  $U(1)^c$ bootstrap at $c=3$ exactly implies that the spectral gap is maximized by a Narain CFT. 
Because we assume the absence of any operators in the  $\ell=1$ sector between $\Delta=1$ and $\Delta=1+\Delta_\phi$, 
we can obtain two-sided bounds on the degeneracy $n_{\Delta=1,\ell=1}$ using the method introduced in Sec.\ref{sec:individual}. In Fig.\ref{fig:c3u1example}, we these bounds on $n_{\Delta=1,\ell=1}$ at derivative order $\Lambda_{max}=23$. Without imposing integrality, $\Delta_g$ at this derivative order is only bounded to be below $\Delta_g=0.7803$. However, at this point the two-sided bounds on the degeneracy do not contain an integer. As the gap $\Delta_g$ is decreased, the upper bound does not change, but the lower-bound slowly decreases until it crosses an integer at approximately $\Delta_g \approx 0.7608$:
\begin{equation}
\Delta_g \le 0.7803\rightarrow \Delta_g  \le 0.7608 \quad (\Lambda_{max}=23) .
\end{equation}
We also compute the lower bound of $n_{\Delta=1,\ell=1}$ for various derivative orders in Fig.\ref{fig:c3u1example}. And in Fig.\ref{fig:c3u1gap}, we show the scalar gap after imposing integrality constraint for derivative orders up to $\Lambda_{max}=51$, and use this to extrapolate to $\Lambda_{\rm max} = \infty$.   At $\Lambda_{max}=51$, the result is \footnote{To get the result at $\Lambda_{max}=51$, one need to be careful about precision. A workable SDPB parameters setting is :$\texttt{precision=1024}$, $\texttt{maxComplementarity=1e+100}$, $\texttt{primalErrorThreshold=1e-40}$, $\texttt{dualErrorThreshold=1e-40}$, $\texttt{dualityGapThreshold=1e-30}$}
\begin{equation}
\Delta_g=0.750746,\quad(\Lambda_{max}=51),
\end{equation}
and we find that the bound extrapolates at $\Lambda_{\rm max}= \infty$ to
\begin{equation}
\Delta_g=0.749 \pm 0.001, \quad(\Lambda_{max}=\infty),
\end{equation}
where the error has been estimated from the variation in different ways to do the extrapolation, see Fig.~\ref{fig:c3u1gap}.
%

We can also compare approximated spectrum coming the extremal functional with the exact $U(1)^{c=3}$ Narain theory. The first few operators from the extremal functional (at $\Lambda_{max}=51$) at $\ell\leq2$ are \footnote{We found multiple spurious operators (operators that don't appear in the $c=3$ Narain theory) in the extremal spectrum. For example, at $\ell=0, \Delta\approx1.366$ there is an operator with degeneracy $n\approx0.256$ from the $\Lambda_{max}=51$ extremal spectrum. We checked that those operator have a small degeneracy, and their degeneracy converges to $0$ when $\Lambda_{max}\to\infty$. We believe these spurious operators will disappear at infinite derivative order. Therefore to avoid confusion, we do not list them in \eqref{eqn:c3spectrum} },
\begin{align}\label{eqn:c3spectrum}
&\ell=0,\quad\Delta=\{0.7507,1.002,2.003,2.749,2.995,4.025,\dots\}\notag\\
&\ell=1,\quad\Delta=\{1,1.7508,2.002,2.992,3.735,3.987,\dots\}\notag\\
&\ell=2,\quad\Delta=\{2.007,2.760,3.006,4.006,4.728,\dots\}
\end{align}
For comparison, the first few operators from $U(1)^{c=3}$ Narain theory are
\begin{align}
&\ell=0,\quad\Delta=\{0.75,1,2,2.75,3,4,\dots\}\notag\\
&\ell=1,\quad\Delta=\{1,1.75,2,3,3.75,4,\dots\}\notag\\
&\ell=2,\quad\Delta=\{2,2.75,3,4,4.75,\dots\}
\end{align}

\section{Eliminating Extremal Solutions}\label{sec:extremal}

Our next application of integrality will be to the set of theories that saturate the Virasoro modular bootstrap bound in the range $1 \le c \le 4$ \cite{Collier:2016cls}.  The extremal functional method already gives an estimate for the degeneracy numbers of these partition functions, 
but here we will go farther and use two-sided bound to rigorously exclude all theories in this range outside of very small numerical windows around the points $c=1,2, \frac{14}{15}$, and $c=4$ discussed previously.

%

\begin{figure}
    \centering
    \subfigure[]{\includegraphics[width=0.49\textwidth]{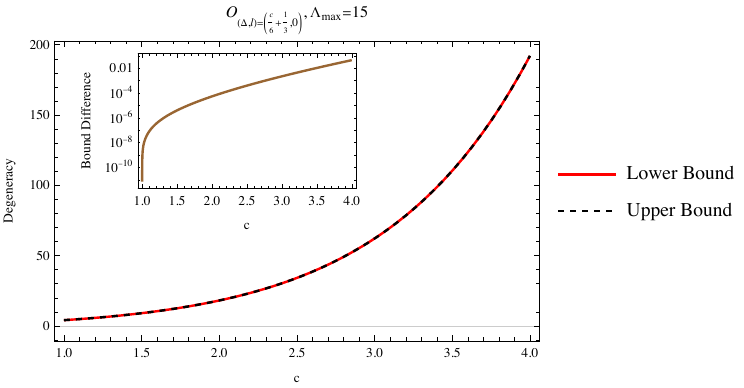}} 
    \subfigure[]{\includegraphics[width=0.49\textwidth]{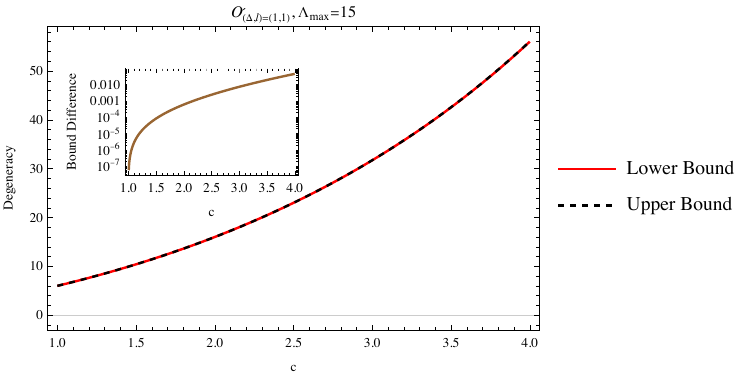}} 
    \subfigure[]{\includegraphics[width=0.49\textwidth]{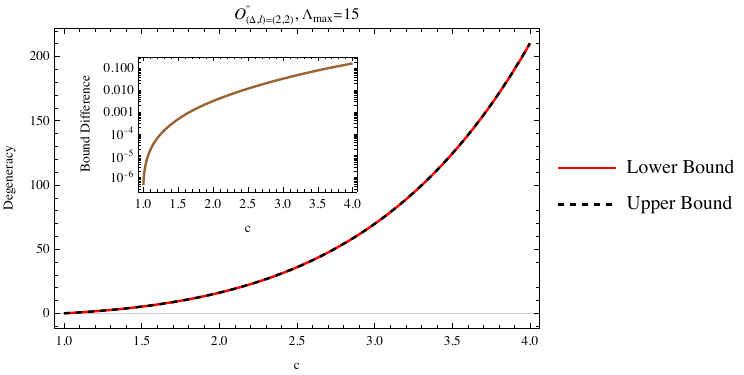}}
    \caption{Panels (a)(b)(c) are the the upper and lower degeneracy bound for operators $\mathcal{O}_{\Delta=\Delta_\phi,\ell=0},\mathcal{O}^{'}_{\Delta=1,\ell=1},\mathcal{O}^{''}_{\Delta=2,\ell=2}$ respectively. The red solid line corresponds to the lower bound while the black dashed line is the upper bound. The inserted plots show the difference between the upper and lower bounds for each operator}
    \label{fig:c1to4bound}
\end{figure}

To begin,  consider the upper and lower bounds of following the scalar operators and two conserved currents operators,
\begin{equation}
\mathcal{O}_{\Delta=\Delta_\phi,\ell=0},\quad\mathcal{O}^{'}_{\Delta=1,\ell=1},\quad\mathcal{O}^{''}_{\Delta=2,\ell=2}
\end{equation}
The extremal functional method applied to theories that saturate the bootstrap bound on the spectral gap tells us the dimensions of the second lowest-dimension operators in the spin $\ell=0,1,$ and $2$ sectors, as observed in equation \eqref{eqn:spectrum}. They are, respectively: 
\begin{equation}
\ell=0,1,2: \quad\Delta_{\text{gap}}=2,1+\Delta_\phi,2+\Delta_\phi , \textrm{ resp.}
\end{equation}
In Fig.\ref{fig:c1to4bound} we show the two-sided bounds for three operators in $\Lambda_{max}=15$. One can see that for three operators, the upper and lower bound is extremely close. First, in  Table.\ref{tb:c1to4theory}, we compare the bounds with the extremal theories at $c=1,2,\frac{14}{5},4$, where one can see that these two sided bounds  fix the degeneracy numbers of these three operators to extremely high precision.

\begin{table}[h!]
\centering
\begin{tabular}{ |c|c|c|c|c|c| } 
 \hline
Theories& &$n_{\frac{c}{6}+\frac{1}{3},0}$ & $n_{1,1}$ & $n_{2,2}$\\ 
\hline
$c=1$&Exact  & 4 & 6 & 0\\ 
&Bound &   {\tiny$(4-5.017*10^{-30},4+5.57*10^{-30})$} & {\tiny$(6-8.96*10^{-28},6+1.16*10^{-29})$} & {\tiny$(-2.666*10^{-28},2.201*10^{-28})$} \\ 
 \hline
$c=2$& Exact & 18 & 16 & 16 \\ 
&Bound & (17.99997,18.00002) & (15.9987,16.0005) & (15.9981,16.0029) \\ 
\hline
$c=\frac{14}{5}$& Exact & 49 & 28 & 54 \\ 
&Bound & (48.9996,49.0007) & (27.9987,28.0035) & (53.9986,54.0209) \\ 
\hline
$c=4$& Exact & 192 & 56 & 210 \\ 
&Bound & (191.986,192.030) & (55.985,56.030) & (209.987,210.154) \\
\hline
\end{tabular}
\caption{The comparison between the degeneracy with the two-sided bounds of the extremal theories. The degeneracy number of the extremal theories are computed in appendix.\ref{app:density} }
\label{tb:c1to4theory}
\end{table}

\begin{figure}
    \centering
   \includegraphics[width=0.7\textwidth]{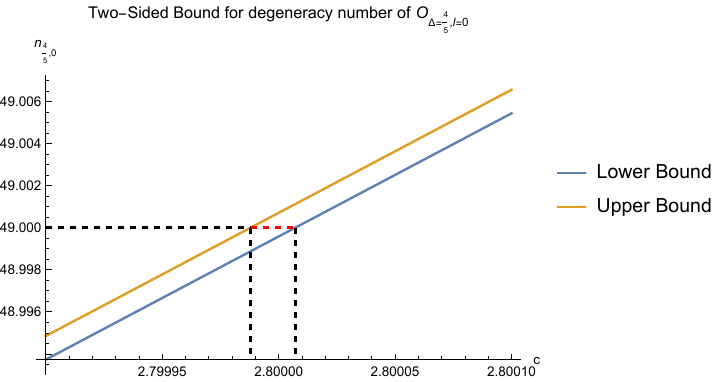}
    \caption{Two sided bound of $n_{\frac{4}{5},0}$ near $c=\frac{14}{5}$. The dashed red region is the only allowed region of central charge within this range}
    \label{fig:c145example}
\end{figure}

Next we use the result in Fig.\ref{fig:c1to4bound} to determine the allowed regions for the central charge. To see how these regions are determine, consider the bound on the degeneracy number of the first scalar operator in the region around $c\sim\frac{14}{5}$. We show the two-sided bound in Fig.\ref{fig:c145example}, where it is clear that the only integer within this range is $n_{\frac{4}{5},0}=49$, and  the allowed region $c\in[c_{\text{min}},c_{\text{max}}]$ is given by when the upper and lower bounds pass through $n_{\frac{4}{5},0}=49$:
\begin{equation}
(N_i,c_{\text{min}},c_{\text{max}})=(49,2.79999,2.800007) .
\end{equation}

\begin{figure}
    \centering
    \subfigure[]{\includegraphics[width=0.45\textwidth]{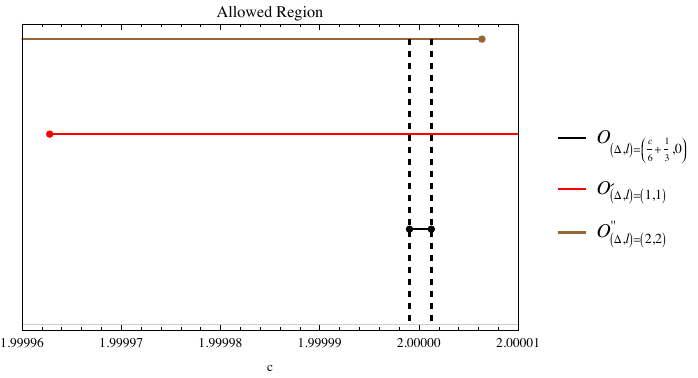}} 
    \subfigure[]{\includegraphics[width=0.43\textwidth]{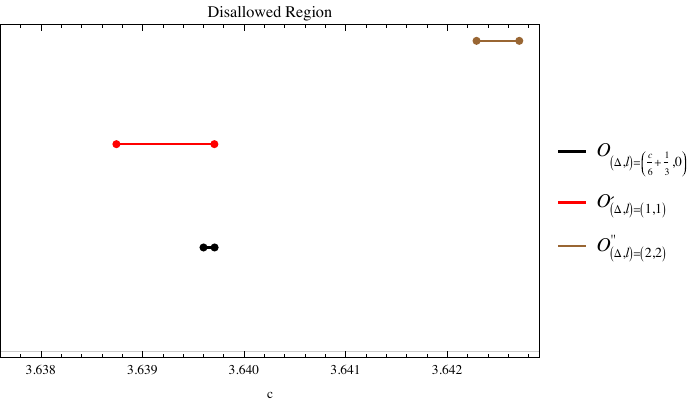}} 
    \caption{Examples of allowed and disallowed region of $c$. \textbf{LHS}: Allowed region of $c$ is between the black dashed line. \textbf{RHS}: No allowed region within this range of $c$}
    \label{fig:regionexample}
\end{figure}

We can repeat this process for any degeneracy number in any range of the central charge.  
  Below we list the first few regions using three different bounds of $\mathcal{O}_{\Delta=\Delta_\phi,\ell=0},\mathcal{O}^{'}_{\Delta=1,\ell=1},\mathcal{O}^{''}_{\Delta=2,\ell=2}$ respectively, at derivative order $\Lambda_{max}=15$:
\begin{align}
&\mathcal{O}_{\Delta=\Delta_\phi,\ell=0},~~N\in[4,192],~~A_1=\{(N_i,c_{\text{min}},c_{\text{max}})\}=\{(4,1,1+3.78\times10^{-9}),\dots\notag\\
&,(5,1.12834993,1.12834995),\dots(192,3.9999,4.00007)\}\notag\\
&\mathcal{O}^{'}_{\Delta=1,\ell=1},~~N\in[6,56],~~A_2=\{(N_i,c_{\text{min}},c_{\text{max}})\}=\{(6,1,1+1.52\times10^{-9})\dots\notag\\
&(7,1.241873,1.241875),\dots(56,3.9999,4.0005)\}\notag\\
&\mathcal{O}^{''}_{\Delta=2,\ell=2},~~N\in[0,210]~~A_3=\{(N_i,c_{\text{min}},c_{\text{max}})\}=\{(0,1,1+3.78\times10^{-9})\dots\notag\\
&(1,1.135886,1.135889)\dots(210,3.9993,4.00006)\}
\end{align}
where $A_1,A_2,A_3$ are allowed regions selected by operators $\mathcal{O}_{\Delta=\Delta_\phi,\ell=0},\mathcal{O}^{'}_{\Delta=1,\ell=1},\mathcal{O}^{''}_{\Delta=2,\ell=2}$. In other words, for the degeneracy number $N$ of, for instance, the operator with spin 0 and dimension $\Delta_\phi$, the set $A_1$ indicates for each possible integer value of $N\in [4, 193]$  the range of $c$s that are consistent with that value of $N$.  In Fig.\ref{fig:regionexample} we show some example plots of  allowed  and disallowed regions. Any consistent value of the central charge $c$ must then lie within one of the allowed regions for all three sets $A_1, A_2, A_3$, since the degeneracy numbers of all three operators must be integers.  The intersection of these three regions $(A_1\cap A_2)\cap A_3$ reduces to a strikingly small set for the allowed ranges of the central charge for the extremal theories:
\begin{equation}
(c_{\text{min}},c_{\text{max}})=\{(1,1+1.52\times10^{-9}),(1.9999,2.000006),(2.7997,2.80002),(3.99986,4.00006)\} .
\end{equation}
These four ranges exactly correspond to the four theories listed in Sec.\ref{sec:bootresult}

\section{Lowering the Spectrual Gap Near $c\sim1$}\label{sec:cnear1}
In Sec.\ref{sec:interval}, we noted that the degeneracy in an interval $\mathcal{F}_\ell(\Delta_a,\Delta_b)$ is a stronger condition than just considering the degeneracy for a specific operator, and in particular it will have two-sided bounds in general.  In this section, we will use this method to slightly lower the spectral gap near $c\sim1$.

\begin{figure}
    \centering
   \includegraphics[width=0.7\textwidth]{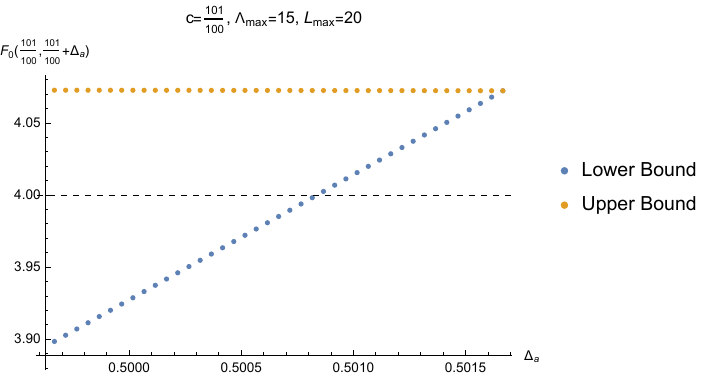}
    \caption{The upper and lower bounds on $\mathcal{F}_0\Big(\Delta_a,\Delta_a+\frac{1}{20}\Big)$ for Virasoro bootstrap, at $c=1+\frac{1}{100},\Lambda_{max}=15$}
    \label{fig:c1p100intervalbound}
\end{figure}

\begin{figure}
    \centering
    \subfigure[]{\includegraphics[width=0.55\textwidth]{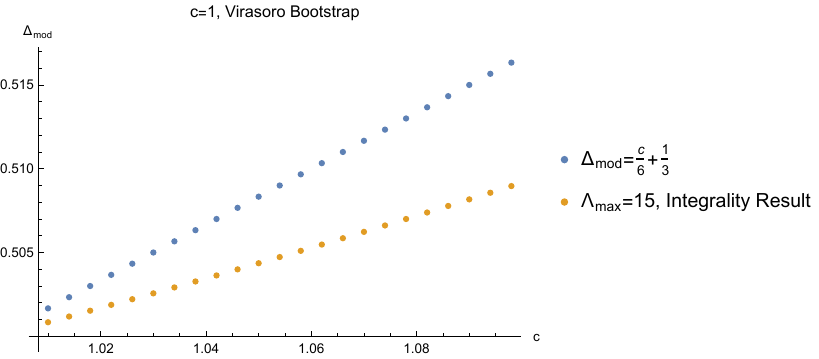}} 
    \subfigure[]{\includegraphics[width=0.37\textwidth]{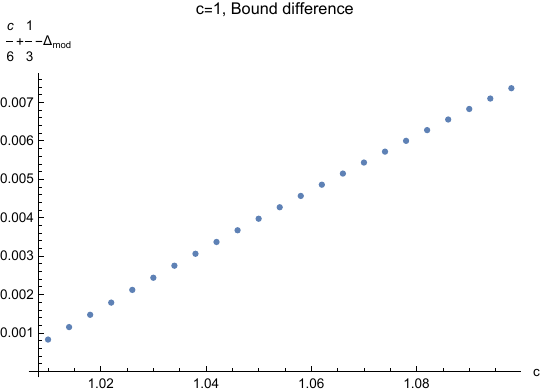}} 
    \caption{(a) The comparison between the integrality result at $\Lambda_{max}=15$ with the convergent bound $\Delta_{\text{mod}}=\frac{c}{6}+\frac{1}{3}$ from \cite{Collier:2016cls} (b) The exact difference between the integrality result at $\Lambda_{max}=15$ and $\Delta_{\text{mod}}=\frac{c}{6}+\frac{1}{3}$}
    \label{fig:cnear1bound}
\end{figure}

We choose two examples,  $c=1.01$ and $c=2$ to illustrate how this works. Define $\Delta_{\text{mod}}$ as the gap of the lowest scaling dimension including all spin sectors. Using only the Virasoro modular bootstrap, the authors in \cite{Collier:2016cls} found that the bound on $\Delta_{\rm mod}$ converged to $\Delta_{\text{mod}}=\frac{c}{6}+\frac{1}{3}$, which in our case is $\frac{301}{600}$ and $\frac{2}{3}$ respectively.
%
We consider the bound on the integrated degeneracy number $\mathcal{F}_0\Big(\Delta_{\text{mod}},\Delta_{\text{mod}}+\frac{1}{20}\Big)$ in the interval $\Delta\in[\Delta_{\text{mod}},\Delta_{\text{mod}}+\frac{1}{20}]$ in the spin $\ell=0$ sector. At $\Lambda_{max}=15$, we  find the following two-sided bounds:
\begin{align}
&c=1+\frac{1}{100},\quad 4.072268\leq\mathcal{F}_0\Big(\frac{301}{600},\frac{301}{600}+\frac{1}{20}\Big)\leq4.072272\Rightarrow\text{Inconsistent!}\notag\\
&c=2,\quad17.944\leq\mathcal{F}_0\Big(\frac{2}{3},\frac{2}{3}+\frac{1}{20}\Big)\leq18.014\Rightarrow\mathcal{F}_0\Big(\frac{2}{3},\frac{2}{3}+\frac{1}{20}\Big)=18.
\end{align} 
From this, we can immediately rule out the case when $c=1+\frac{1}{100}$ with $\Delta_{\text{mod}}=\frac{301}{600}$. 
By contrast, at $c=2$ the extremal theory with $\Delta_{\text{mod}}=\frac{2}{3}$ corresponds to the $SU(3)_1~WZW$ model, and passes the consistency check as expected.  
At $c=1.01$, we can now go farther and improve the bound on $\Delta_{mod}$. In Fig.\ref{fig:c1p100intervalbound}, we plot the upper and lower bound of $\mathcal{F}_0\Big(\Delta_a,\Delta_a+\frac{1}{20}\Big)$ as a function of $\Delta_a$ at $\Lambda_{max}=15$. We obtain an improved value for the bound on $\Delta_{\text{mod}}$ from the point where $\mathcal{F}_0\Big(\Delta_a,\Delta_a+\frac{1}{20}\Big)$ first includes $4$. This occurs at the value $\Delta_{a} \approx 0.5008$, whereas the previous bound was $\Delta_{\rm mod} = \frac{301}{600} \approx 0.5017$.  
In Fig.\ref{fig:cnear1bound} we show the result more generally for a range of $c$ near $c\sim 1$ at $\Lambda_{max}=15$.\footnote{For the case where $c$ is very large, this method gives significantly less improvement in the bound; for example, at $c=3$ we found that at $\Lambda_{max}=15$, the decrease in the spectral gap is less than $10^{-5}$.}


\newpage
\begin{center}
\subsection*{Acknowledgments}
\end{center}
We thank David Simmons-Duffin and Yuan Xin for helpful conversations, and Yuan Xin for comments on the earlier draft. ALF and WL are supported by the US Department of
Energy Office of Science under Award Number DE-SC0015845, and the Simons Collaboration on the Non-Perturbative Bootstrap. ALF thanks the Aspen Center of Physics for hospitality as this work was completed.

\newpage

\appendix
\section{Partition function for extremal theories}\label{app:density}
In the section we list the partition function extremal theories with $c=1,2,\frac{15}{4},4$ in Sec.\ref{sec:extremal}. The partition function for extremal theory in $c=1,2,4$ can be found in \cite{Collier:2016cls}. The partition function $(G_2)_1$ $WZW$ model at $c=\frac{14}{5}$ can be found in \cite{Rayhaun:2023pgc},
\begin{align}
&(q\bar{q})^{-\frac{1}{24}}|\eta(\tau)|^2\Big[Z_{\text{ext}}^{c=1}(\tau,\bar{\tau})-\chi_{vac}(\tau)\bar{\chi}_{vac}(\bar{\tau})\Big]=4(q\bar{q})^{\frac{1}{4}}+3(q+\bar{q})+3q\bar{q}+\dots\notag\\
&(q\bar{q})^{\frac{1}{24}}|\eta(\tau)|^2\Big[Z_{\text{ext}}^{c=2}(\tau,\bar{\tau})-\chi_{vac}(\tau)\bar{\chi}_{vac}(\bar{\tau})\Big]=18(q\bar{q})^{\frac{1}{3}}+8(q+\bar{q})+\bar{q}^{\frac{4}{3}}q^{\frac{1}{3}}+8(q^2+\bar{q}^2)+48q\bar{q}+\dots\notag\\
&(q\bar{q})^{\frac{3}{40}}|\eta(\tau)|^2\Big[Z_{\text{ext}}^{c=\frac{14}{5}}(\tau,\bar{\tau})-\chi_{vac}(\tau)\bar{\chi}_{vac}(\bar{\tau})\Big]=49(q\bar{q})^{\frac{2}{5}}+14(q+\bar{q})+27(q^2+\bar{q}^2)+168q\bar{q}+\dots\notag\\
&(q\bar{q})^{\frac{1}{8}}|\eta(\tau)|^2\Big[Z_{\text{ext}}^{c=4}(\tau,\bar{\tau})-\chi_{vac}(\tau)\bar{\chi}_{vac}(\bar{\tau})\Big]=192(q\bar{q})^{\frac{1}{2}}+28(q+\bar{q})+105(q^2+\bar{q}^2)+728q\bar{q}+\dots
\end{align}
Expanding in the first few order in $q,\bar{q}$, we can extract the degeneracy for operators $\mathcal{O}_{\Delta=\frac{c}{6}+\frac{1}{3},\ell=0},\mathcal{O}^{'}_{\Delta=1,\ell=1},\mathcal{O}^{''}_{\Delta=2,\ell=2}$. Explicitly,
\begin{equation}
(q\bar{q})^{\frac{c-1}{12}}|\eta(\tau)|^2\Big[Z_{\text{ext}}(\tau,\bar{\tau})-\chi_{vac}(\tau)\bar{\chi}_{vac}(\bar{\tau})\Big]=n_{\frac{c}{6}+\frac{1}{3},0}(q\bar{q})^{\frac{c}{12}+\frac{1}{6}}+\frac{n_{1,1}}{2}(q+\bar{q})+\frac{n_{2,2}}{2}(q^2+\bar{q}^2)+\dots
\end{equation}
Thus for these four extremal theories, we have,
\begin{align}
&c=1:\quad n_{\frac{1}{2},0}=4,\quad n_{1,1}=6,\quad n_{2,2}=0\notag\\
&c=2:\quad n_{\frac{2}{3},0}=18,\quad n_{1,1}=16,\quad n_{2,2}=16\notag\\
&c=\frac{14}{5}:\quad n_{\frac{4}{5},0}=49,\quad n_{1,1}=28,\quad n_{2,2}=54\notag\\
&c=4:\quad n_{1,0}=192,\quad n_{1,1}=56,\quad n_{2,2}=210
\end{align}

\bibliographystyle{JHEP}
\bibliography{refs}

\end{document}